\documentclass{vgtc}                          

\graphicspath{{figures/}{pictures/}{images/}{./}} 

\usepackage{times}                     

\usepackage{tabu}                      
\usepackage{booktabs}                  
\usepackage{lipsum}                    
\usepackage{mwe} 
\usepackage{amsmath}
\usepackage{booktabs}
\usepackage{tabularx}
\usepackage{float}

\usepackage{mathptmx}                  

\onlineid{6596}

\vgtccategory{Research}

\vgtcinsertpkg

\title{Evaluating Just Noticeable Differences in Layered Opacity Visualizations}

\author{Caterina Ponti\thanks{e-mail: cponti@dons.usfca.edu}\\ %
        \scriptsize University of San Francisco %
\and Shano Liang\thanks{e-mail: sliang1@wpi.edu}\\ %
     \scriptsize Worcester Polytechnic Institute %
\and Lane Harrison\thanks{e-mail: ltharrison@wpi.edu}\\ %
     \scriptsize Worcester Polytechnic Institute %
\and Alark Joshi\thanks{e-mail: apjoshi@usfca.edu}\\ %
     \scriptsize University of San Francisco}
     
\teaser{
  \centering
  \includegraphics[width=\linewidth]{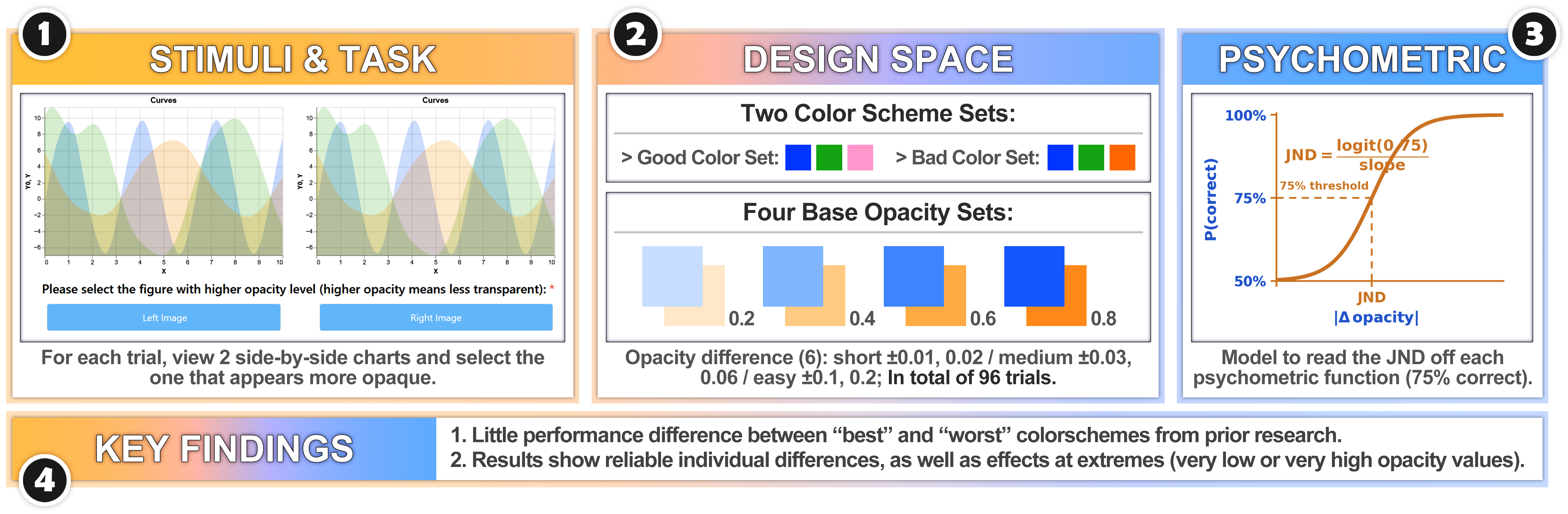}
  \caption{Overview of our study. We measure just-noticeable differences (JNDs) in layered opacity charts. 
  (1) On each trial, participants view two side-by-side charts and select the one that appears more opaque; 
  (2) Stimuli span a controlled design of four base opacity levels ($0.2–0.8$), six opacity differences, and the two color schemes drawn from Meng et al. \cite{meng2025seeing}: a recommended (``\textit{good}'') scheme and a non-recommended (``\textit{bad}'') scheme, for 96 trials per participant; 
  (3) We fit a Bayesian logistic mixed-effects model to the responses and read the JND off each psychometric function as the opacity difference giving 75\% correct performance; 
  (4) Results suggest little difference between colorschemes, but larger effects for individual differences and variance when comparisons approach extreme values (0.0 or 1.0).}
  \label{fig:teaser}
}

\abstract{
Opacity is a widely used channel in data visualization, but it remains less well understood compared to channels such as color, length, size, etc.
Recent work from Meng et al.~\cite{meng2025seeing}, investigated the impact of opacity across competing color schemes, finding that certain color schemes were associated with better participant accuracy.
We examine these effects further in a controlled two-alternative forced choice setup, to determine whether opacity differences are truly equal across possible opacity comparison ranges. 
In a within-subjects study with 96 trials, including two competing color schemes (\textit{best} and \textit{worst} from Meng et al.~\cite{meng2025seeing}) and 48 opacity pairs, we find little differences between color schemes, but larger individual differences in accuracy. Further, results show stable performance in middle opacity ranges, with more errors occurring when comparing extreme values. We discuss potential implications for design guidelines and further study, and make our study materials, analysis scripts, and data available at \url{https://osf.io/zv9dx/overview?view_only=38d03cea1b3d42788593e3e6b1016cfd}.
} 


\begin{document}


\firstsection{Introduction}

\maketitle
Understanding how people perceive visual differences is essential for designing effective visualizations. In this study, we investigate how accurately people can detect \textit{differences in opacity when interpreting layered opacity charts}. Specifically, we use a two-alternative forced choice task and estimate JNDs across opacity levels and color schemes to estimate the change in opacity where observers can reliably perceive differences. 

Building on existing literature on perception comparison in visualizations~\cite{harrison2014ranking,kay2015beyond,rensink2010perception}, we focused on layered opacity chart comparisons and examined participants' ability to detect opacity differences using a fixed set of comparison stimuli. We tested four baseline opacity levels (0.2, 0.4, 0.6, and 0.8) and compared each to stimuli with opacity differences of $\pm$0.01 and $\pm$0.02 (short), $\pm$0.03 and $\pm$0.06 (medium), and $\pm$0.1 and $\pm$0.2 (long). Trials were presented in randomized order, and every participant completed trials in both the ``good'' and ``bad'' color conditions, which we adapted from Meng et al. \cite{meng2025seeing}.

We analyzed how visual opacity discrimination accuracy varies as a function of the magnitude of the opacity difference between layered opacity plots, the baseline opacity level of stimuli, and the experimental condition (good vs. bad).
Collected trial data were modeled using logistic regression with the base opacity level and color-scheme serving as the main factors, allowing for variation in parameters between participants to test for potential individual differences.

Results suggest that color-scheme had relatively little impact on performance in terms of estimated JNDs. However, results did show more variation between participants, with some detecting even small differences reliably. Comparisons at the extremes (e.g. near 0.0 or 1.0) also resulted in variance in performance. We discuss potential implications of these findings in terms of design and future studies. 

\section{Related Work}

Many prior studies in visualization evaluate how well people perceive differences or estimate values based on carefully controlled visual stimuli, often reflecting how data is encoded in practice. Previous studies have explored how changes in visual attributes (correlation~\cite{rensink2010perception, Strain2023TheEOH,Zeng2021ARAL}, color~\cite{Tseng2023MeasuringCPK, Zeng2021ARAL} and opacity~\cite{Strain2024EffectsOPI, meng2025seeing}) affect people's performance. 

\subsection{Scatterplots, Correlation, and JNDs}

Rensink et al.~\cite{rensink2010perception} evaluated the visual perception of correlation in scatterplots by measuring the observers’ sensitivity to differences in correlation. Precision was measured using the Just Noticeable Difference (JND) (in the context of correlation perception) that is defined as the minimum difference between two side-by-side correlations that can be discriminated 75 percent of the time. Rensink et al. showed that the perception of positive correlation in scatterplots can be modeled using Weber’s Law~\cite{ekman1959weber}. Subsequently, Harrison et al.~\cite{harrison2014ranking} used Weber’s Law to evaluate the effectiveness of commonly used visualizations, demonstrating that nine commonly used visualizations follow Weber’s law. 
The Just Noticeable Difference (JND) is a key component in this evaluation, as it defines the smallest detectable change in correlation perception, helping to establish perceptual thresholds. 
Kay et al.~\cite{kay2015beyond} reviewed the generality of Weber’s Law and found that a log-linear model provides a better description of the relationship between JNDs and objective correlation strength. Their Bayesian linear mixed-effects model accounts for individual variability. It suggests that correlation perception may follow a more complex, non-linear pattern than previously assumed.
Elliott and Rensink \cite{Elliott2015InterferenceITC} further showed that overlapping visual populations can interfere with the perception of correlation. Their findings suggest that perceptual thresholds may change when multiple visual layers compete for attention. 


Beyond correlation estimation, Micallef et al. \cite{Micallef2017TowardsPOB} demonstrated that scatterplot design parameters such as point size, transparency, and sampling influence perceptual accuracy, highlighting the importance of accounting for visual encoding choices when modeling perceptual thresholds.

\subsection{Color and Opacity Studies}
Lu et al.~\cite{lu2024color} addressed challenges in transparency perception. When color overlaps with transparency, they blend, creating ambiguous color mappings and introducing false colors. To solve this issue, Lu et al. introduced a color-name-aware (CNA) optimization framework that produced consistent color choices and transparency levels while maintaining a clear visual distinction between all parts of the graph. They also showed that their method works for histograms and other visualizations like parallel coordinate plots and Venn diagrams. Meng et al.~\cite{meng2025seeing} investigated how color hue and opacity affect depth perception in semi-transparent visualizations. Their findings underscore that both color and opacity interact to shape perceptual accuracy in layered visual representations, further motivating the need to model perceptual thresholds such as JNDs in visualization design, especially when looking at the joint effect of opacity and color. Specifically, they identified 4 key predictors of depth order perception: hue front, hue back, opacity back, and opacity front. They also found that participants made \textit{fewer errors when certain colors appeared in specific layer positions}: blue, green, and orange had fewer errors when placed in the front layer, and orange and pink did so when placed in the back layer. However, their study did not examine how participants perform when comparing side-by-side layered opacity charts that differ in opacity levels. Building on these findings, our experiment draws primarily from the methodologies and insights of Kay et al.~\cite{kay2015beyond} and Meng et al.~\cite{meng2025seeing} to investigate perceptual threshold in layered opacity visualizations.

\section{Methods}

\begin{figure}[ht]
\centering
\includegraphics[width=\columnwidth]{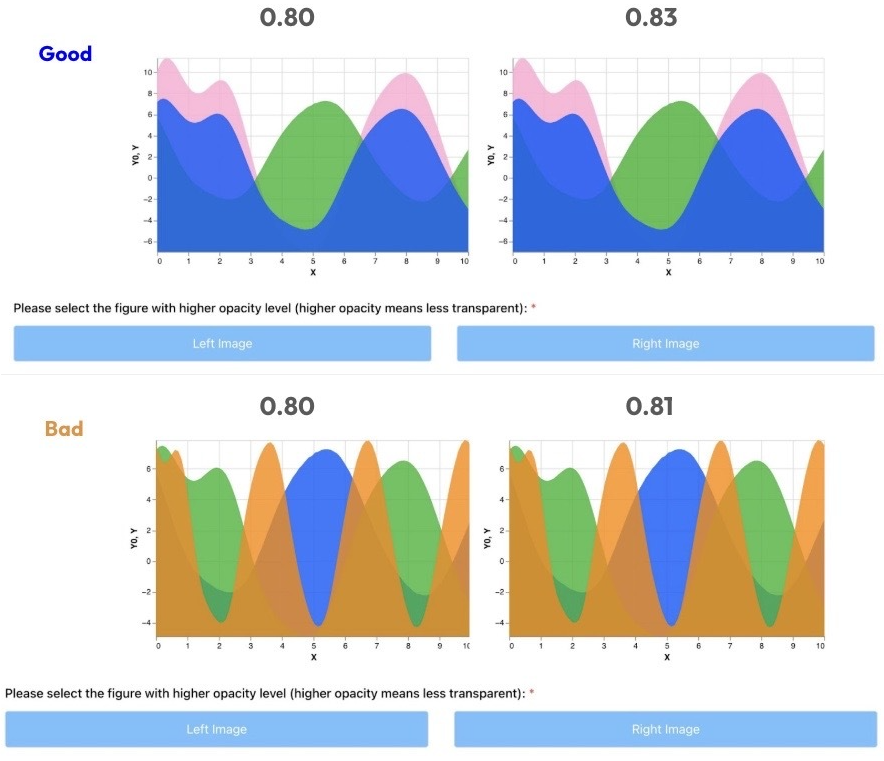}
\caption{This figure shows two example stimuli with the interface seen by the participants. The top row showing the ``good'' palette showing the 0.8 stimuli with the 0.83 stimuli (JND distance of 0.3). The bottom row shows the ``bad'' palette with the 0.8 stimuli and the 0.81 stimuli. The stimuli in the bottom row are indistinguishable according to the results of our study. You can see a demo of the study at  \protect\url{https://opacityjnd.web.app}.}
\label{fig:stimuli}
\end{figure}

The experiment investigated opacity discrimination under two color conditions derived from color recommendations reported in previous work~\cite{meng2025seeing}. 
The recommended color combination consisted of blue and green elements in the front layer and orange and pink elements\footnote{The \textit{HSL} (Hue, Saturation, and Lightness) values of these colors are $(219,100,50)$ for blue, $(119,58,44)$ for green, $(332,100,83)$ for pink, and $(31,100,51)$ for orange.} in the back layer. Previous findings showed that blue, green, and orange (``good palette'') produced lower error rates when displayed in the front layer, whereas orange and pink produced lower error rates when displayed in the back layer. The non-recommended color combination consisted of orange elements in the front layer and blue and green elements in the back layer (``bad palette''). This combination was associated with higher error rates in prior work and served as a comparison condition. The goal in including these was to determine whether error rates across ``good'' and ``bad'' palettes were associated with particular opacity levels and differences, as well as to examine individual performance.
Stimuli were rendered using Vega with its default HTML5 Canvas renderer against a white background (\#FFFFFF; RGB 255, 255, 255). 
Opacity was applied using the browser's default Canvas 2D source-over compositing in gamma-encoded sRGB. 

\begin{figure}
\centering
\includegraphics[width=\columnwidth]{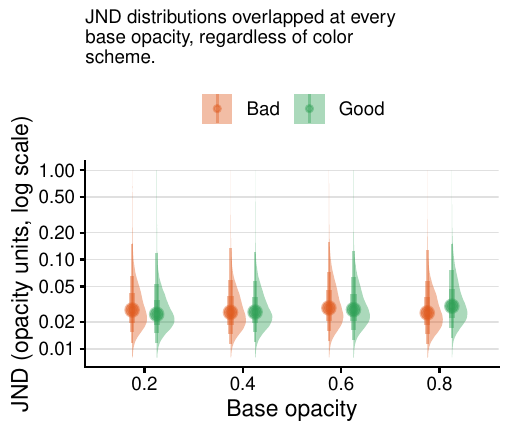}
\caption{Full posterior JND distributions by color condition and base opacity (log scale). Distributions largely overlap between the recommended (good) and non-recommended (bad) conditions at every base opacity level.}
\label{fig:jnd_dist}
\end{figure}

\begin{figure}
\centering
\includegraphics[width=\columnwidth]{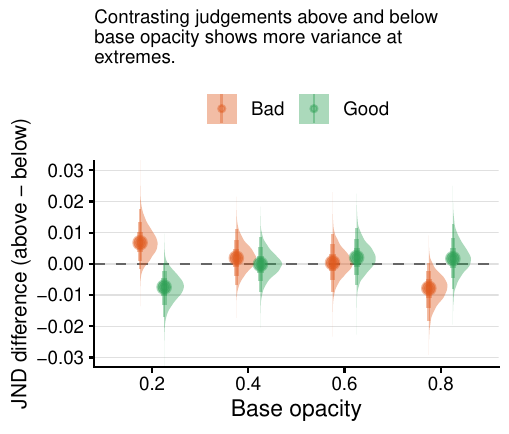}
\caption{Within-condition direction contrast (above $-$ below JND) by color condition and base opacity. Positive values indicate the "below" direction was easier to detect (lower JND); negative values indicate "above" was easier. Bad and good conditions sit on opposite sides of zero at base opacity 0.2 and flip at base opacity 0.8.}
\label{fig:within_condition_contrast}
\end{figure}

Stimuli were presented at four base opacity levels (0.2, 0.4, 0.6, and 0.8) in both color conditions. On each trial, two images differing in opacity were displayed side by side with the participant asked to choose the figure with the higher opacity, as shown in Figure~\ref{fig:stimuli}. Figure~\ref{fig:stimuli} shows the ``good'' palette combination in the top row with two stimuli (with opacity 0.80 and 0.83 (+0.03 = JND distance from the original stimuli)) and two stimuli (with opacity 0.80 and 0.81) with the ``bad'' palette in the bottom row. The opacity difference varied across trials. For each opacity, the comparison stimuli distances ranged from short ($\pm$0.01) to long ($\pm$0.2).


Participants completed both color conditions in a mixed design. Each participant performed 96 trials corresponding to all combinations of four base opacity levels and two color conditions (12 stimuli distances * 2 color conditions * 4 opacities). On every trial, participants viewed two images and indicated which image appeared to have \textit{greater opacity} by selecting either the left or the right image. Response accuracy was recorded for each trial.

\section{Results}

\subsection{Participants}
A total of 30 participants were recruited through Prolific to complete the online opacity experiment. Initial analyses revealed potential confusion among participants who flipped responses. As these also fell more than two standard deviations below mean performance, they were excluded. Mean accuracy across participants was 75.3\% (SD = 20.9\%), resulting in an exclusion threshold of 33.5\%. Three participants were excluded, with a final sample of 27 participants.
Participant reported demographics including 18 men (66.7\%), 8 women (29.6\%), and 1 non-binary (3.7\%). Participants ranged in age from 18 to 64 years. With respect to education, 13 participants (48.1\%) reported holding a bachelor's degree or equivalent, 8 (29.6\%) reported a high school diploma or equivalent as their highest level of education, and 6 (22.2\%) reported holding a master's degree or equivalent. 

\subsection{Statistical Analysis}
Accuracy was modeled using a Bayesian logistic mixed-effects model\footnote{Data and analysis scripts available at: \url{https://osf.io/zv9dx/overview?view_only=38d03cea1b3d42788593e3e6b1016cfd}}. The probability of a correct response was predicted by a separate slope for each combination of color condition and base opacity (i.e., an interaction with no main-effect or intercept terms), so that predicted accuracy is exactly chance when the opacity difference is zero. By-participant random effects were included for the intercept, the opacity-difference slope, the color-condition effect, and their interaction, to account for individual differences in both overall sensitivity and in how sensitivity varied by color condition:

\begin{multline*}
P(\text{correct}) \sim |\Delta \text{opacity}| : \text{condition} : \text{base opacity} - 1 \\
+ \left(|\Delta \text{opacity}| \times \text{condition} \mid \text{participant}\right)
\end{multline*}

Just Noticeable Difference (JND) was defined as the \textit{opacity difference required to achieve 75\% correct performance}. At the population level, where the fixed effects have no intercept, the JND is simply the opacity difference at which the fitted slope reaches the 75\%-correct criterion:

\[
\mathrm{JND}
=
\frac{\operatorname{logit}(0.75)}{\text{slope}}
\]

Larger JND values indicate lower sensitivity, meaning a greater opacity difference is needed to achieve the same level of performance.
Larger opacity differences were associated with higher probabilities of correct responses across all eight conditions $\times$ base-opacity combinations. 

\subsection{Just-Noticeable Difference Estimates}

JND values were defined as the opacity difference corresponding to 75\% predicted accuracy. 
Posterior median JND estimates and 95\% credible intervals are shown, in full distributional form, in Figure~\ref{fig:jnd_dist}. 
At base opacity 0.2, the bad condition yielded a somewhat larger median JND (0.0271, 95\% CrI [0.0107, 0.166]) than the good condition (0.0243, 95\% CrI [0.0107, 0.126]), though the credible intervals overlap substantially and the posterior probability that bad exceeded good was $P(\text{bad}>\text{good})=0.774$. This pattern was not consistent across the opacity range: contrasts at base opacities 0.4 and 0.6 were inconclusive, and at base opacity 0.8 the point estimate nominally reversed (good: 0.0300 vs.\ bad: 0.0251), though this contrast was also not clearly credible ($P(\text{bad}>\text{good})=0.096$). Overall, JND did not vary monotonically with base opacity, and pooling across the full opacity range showed little evidence of a consistent difference between color conditions.

\begin{figure*}[t]
\centering
\includegraphics[width=\textwidth]{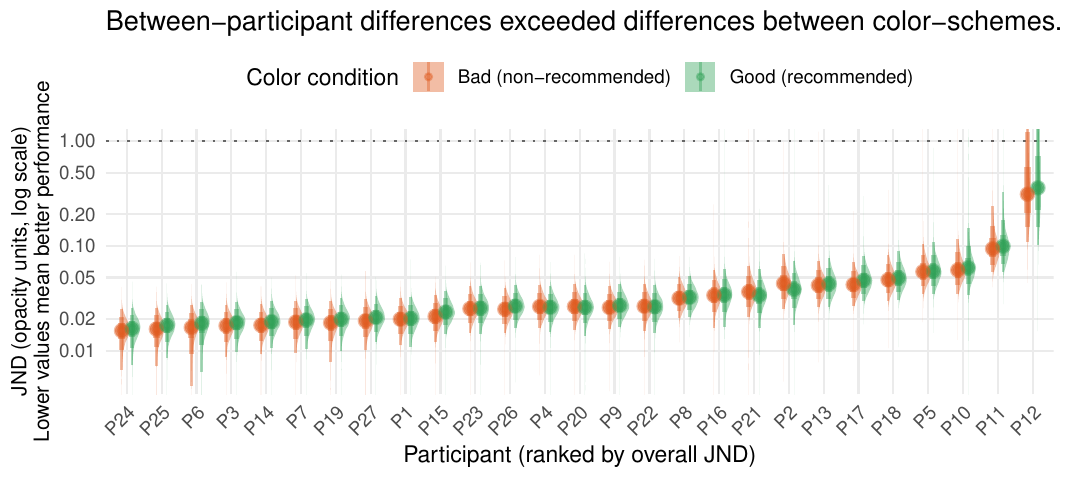}
\caption{Between-participant differences in overall sensitivity. Each participant's JND (by color condition), averaged across base opacity, ranked from most to least sensitive. Between-participant spread is substantially larger than the difference between color conditions within any given participant.}
\label{fig:participant_rank}
\end{figure*}

\subsection{Overall Color-Scheme Effects}
Prior work on layered opacity visualizations has devoted substantial attention to how color-scheme choices interact with opacity to shape perceptual accuracy~\cite{meng2025seeing}. Given this emphasis, a natural first question is whether the recommended (``good'') and non-recommended (``bad'') color schemes produced systematically different JNDs once results are pooled across the full range of base opacities and comparison directions. As shown in Figure~\ref{fig:jnd_dist}, the full posterior JND distributions for the two color schemes overlap substantially at every base opacity level, and none of the four base-opacity contrasts reached conventional credibility. 
We therefore find \textit{no reliable overall effect of color scheme on opacity discrimination in this task}.

\subsection{Individual Differences}
In contrast to the modest overall effect of color scheme, participants varied more substantially from one another in their overall ability to discriminate opacity differences (Figure~\ref{fig:participant_rank}). Some participants performed extremely well, reliably discriminating opacity differences of roughly 1 to 1.5\%, while others required a difference several times larger to perform at the same level, with a broad middle range of participants falling in between. One participant (P12) stands out with a JND far outside the range of the rest of the sample and a correspondingly wide credible interval, consistent with near-chance performance on a subset of trials rather than genuine low perceptual sensitivity.

For example, at a base opacity of 0.80 in the recommended (``good'') color condition-- the best-performing participant could reliably tell 0.80 apart from a comparison opacity of just 0.82. In comparison, the observed worst-performing participant (excluding P12, given the near-chance performance) would need the comparison pushed all the way to 0.93 before reliably seeing differences between stimuli.


\subsection{Differentiating Above/Below Trials}
To test whether the effect of color condition depended on whether the comparison stimulus was more or less opaque than the base (``above'' vs.\ ``below''), trials were split by this direction, and the model was refit with this additional factor. Figure~\ref{fig:within_condition_contrast} shows the resulting within-condition contrast (above $-$ below JND) for each color condition and base opacity.

This split revealed a pattern that the pooled results in Figure~\ref{fig:jnd_dist} could not show: at the extremes of the tested opacity range (0.2 and 0.8), the two color schemes diverged in opposite directions specifically for comparison stimuli that were more opaque than the base. At base opacity 0.2, the ``bad'' scheme yielded a higher JND (worse performance) than the ``good'' scheme in this direction; at base opacity 0.8, this pattern reversed, with the good scheme instead yielding the higher JND. Contrasts in the opposite direction (comparison less opaque than the base) and at the two intermediate base opacities did not show this pattern. Rather than a uniform or monotonic difference between schemes, any true effect of color scheme, therefore, appears to be confined to specific combinations of comparison direction and base opacity, consistent with the absence of a simple overall effect reported above.

\section{Discussion}
Results suggest that opacity discrimination was consistent for individual participants, regardless of the ``good'' or ``bad'' color scheme. 
This invites further investigation on what actually drives error rates in visualization tasks involving opacity, such as the depth-order judgments studied by Meng et al.~\cite{meng2025seeing} or the scatterplot design parameters (point size, transparency, sampling) studied by Micallef et al.~\cite{Micallef2017TowardsPOB}.
One possibility is that these tasks differ in their underlying mechanisms, judging depth order or estimating correlation, versus our task which required discrimination between opacity values. These tasks may recruit distinct perceptual, cognitive, or decision-making processes despite all involving judgements of opacity at some level.
Future studies might explore the relationship between these tasks in a single study so that within-participant comparisons are possible across tasks and visualization types.

While color scheme did not have a main effect, results did reveal two other effects to consider.
Most notable is the effect of individual differences. 
Some participants were remarkably adept at the task, correctly discriminating opacity when it differed by about 1-1.5\%.
However, some participants needed a 5\% or more difference to reliably be able to distinguish between opacity differences. 
While further investigation is warranted (e.g., see limitations below on visualization types and tasks), this may point towards practical guidelines of designing opacity difference ranges in 5\% or greater ``steps'', to help ensure participants can reliably distinguish differences.

Results also showed modest differences at the extremes of the ranges tested, with error rates of the ``bad'' scheme improving at higher ranges and error rates of the ``good'' scheme improving in lower ranges. Curiously, the middle opacity values seem unaffected. 
We do not believe this points towards a potential future recommendation as much as a need for further studies. We recommend testing more color schemes and increased base opacity ranges (e.g., getting close to the extremes) to determine whether there are any effects of color scheme and performance at extreme ranges.

Also worth noting are limitations. Primarily, we tested a single visualization type, layered area charts, due to their common use of opacity. However, many other visualizations often use the opacity channel to help deal with various forms of overlap and overplotting. For the present work, our goal was to evaluate the overall methodology and test for effects across opacity values and individuals. Another limitation is the task itself. Discrimination is not the same as estimating values and should not be taken as having a direct connection to higher-level visualization task performance. At the same time, discrimination has been used to investigate visualization stimuli in controlled settings for some time now, often revealing useful insights into the relationship between perceptual performance and data visualizations. 

The present study also lays groundwork for several potential future investigations.
Future work might expand on the modeling approach, for example by calculating differences not based on opacity differences but based on perceptual color differences for each stimulus pair. Future studies should also include more in-depth training trials, which may help with the formation of their strategies. Participants could even be asked to describe their comparison strategies, which might reveal cases where participants appeared to reverse judgments, as well as important insights into the underlying reasons behind the performance differences between participants.

\bibliographystyle{abbrv-doi}

\bibliography{template}
\end{document}